\documentclass[11pt,letterpaper]{article}

\usepackage{jheppub}
\usepackage{color}
\usepackage{graphicx}
\usepackage{wrapfig,enumerate,slashed}

\usepackage{footmisc}
\usepackage{amsmath}
\usepackage{wasysym} 
\usepackage{graphicx}
\usepackage{color}
\usepackage{comment}
\usepackage{hyperref}

\newcommand{\be}{\begin{eqnarray}}
\newcommand{\ee}{\end{eqnarray}}

\newcommand{\bee}{\begin{eqnarray}}
\newcommand{\eee}{\end{eqnarray}}
\newcommand{\beeq}{\begin{equation}}
\newcommand{\eeeq}{\end{equation}}

\renewcommand{\vec}{\bf}

\DeclareRobustCommand{\Sec}[1]{Sec.~\ref{#1}}

\DeclareRobustCommand{\Fig}[1]{Fig.~\ref{#1}}

\DeclareRobustCommand{\Eq}[1]{Eq.~(\ref{#1})}

\begin{document}

\title{Combining LEP and LHC to bound the Higgs Width}

\author[a]{Christoph Englert,}
\author[b]{Matthew McCullough,}
\author[c]{and Michael Spannowsky}

\affiliation[a]{SUPA, School of Physics and Astronomy, University of Glasgow, Glasgow, G12 8QQ, UK}
\affiliation[b]{Theory Division, CERN, 1211 Geneva 23, Switzerland}
\affiliation[c]{Institute for Particle Physics Phenomenology, Department of Physics, Durham University, DH1 3LE, UK}

\emailAdd{christoph.englert@glasgow.ac.uk}
\emailAdd{matthew.mccullough@cern.ch}
\emailAdd{michael.spannowsky@durham.ac.uk}

\abstract{
The correlation of on- and off-shell Higgs boson production at the LHC in $gg \to h^* \to ZZ$ to bound the Higgs width, under specific model-dependent assumptions, has recently received a lot of attention.  As off-shell cross section measurements in this channel suffer from a small signal yield, large backgrounds, and theoretical uncertainties, we propose an alternative complementary constraint which is only possible through the combination of LEP and LHC measurements.  Previous precision electroweak measurements at LEP allow for the determination of indirect constraints on Higgs couplings to vector bosons by considering one-loop processes involving virtual Higgs exchange.  As the Higgs is off-shell in these diagrams we venture that LEP can be interpreted as an off-shell `Higgs Factory'.  By combining these LEP constraints with current LHC 8 TeV Higgs measurements a stronger limit on the Higgs width can be achieved than with LHC data alone for models with rescaled Higgs couplings.  Looking to the future, avoiding ambiguities arising due to new physics modifications of the $hGG$ coupling, a theoretically more robust constraint can be achieved by correlating LEP measurements with WBF Higgs production followed by Higgs decays to $WW$ and $ZZ$.  This method for indirectly constraining the Higgs width is very effective for specific BSM scenarios and is highly complementary to other proposed methods. The limits we obtain particularly highlight the power of a concrete LEP+LHC combination, not only limited to Higgs width measurements.
}

\preprint{DCPT/15/26, IPPP/15/13, CERN-PH-TH-2015-077}
\maketitle

\section{Introduction}
\label{sec:intro}
The discovery of the Higgs boson \cite{orig,Chatrchyan:2012ufa,Aad:2012tfa} marked a new era of exploration in fundamental physics.  Ideally one would like to be able to extract all of the properties of the Higgs, such as the mass and individual decay widths, as well as detailed information on the coupling magnitudes and Lorentz structures.  In practice such an extensive wish list cannot be met with direct measurements alone, and varying degrees of theoretical assumptions must be imposed in order to map from measurement to Lagrangian.

In this work we will consider the total Higgs decay width, which is a crucial parameter for many scenarios beyond the Standard Model (SM). The total Higgs width has received considerable experimental and theoretical attention recently \cite{atlas,Khachatryan:2014iha,ChrisandMichael,offshella,offshellb,Offshell,ilc} after a recent proposal for correlating on- and off-shell Higgs production at the LHC~\cite{Caola:2013yja,Kauer:2013qba,ciaran}. In this paper we pursue a different strategy, which combines the off-shell Higgs information gathered at the Large Electron Positron Collider (LEP) with LHC Higgs measurements. It is important to realise that, although the discovered Higgs falls outside the {\it kinematic} coverage of LEP, the high precision results of LEP still provide seminal and complementary information to current and future Higgs physics analyses, especially now that it has been established that $m_h\simeq 125$~GeV.

The implications of the Higgs discovery for the combined electroweak parameter fit was analysed in \cite{Baak:2014ora}. Our work takes a different approach and concretely uses the Higgs coupling information determined from the LEP results in correlation with LHC Higgs measurements to constrain free parameters in entire classes of models. While our discussion will be focused on the Higgs width, it is important to stress that this strategy is applicable in a much broader context. The limits we obtain particularly highlight the power of a concrete LEP+LHC combination.

We organise this work as follows: First we review recent attempts to set limits on the total Higgs width at the LHC in Sec.~\ref{sec:hwidth} and argue further in Sec.~\ref{sec:LEP} that in some circumstances LEP can be considered a superior off-shell Higgs constraint.  In Sec.~\ref{sec:currentcombo} we establish this quantitatively by combining LEP and current LHC 8 TeV results to set a constraint on the total Higgs width in the spirit of Refs.~\cite{Caola:2013yja,atlas,Khachatryan:2014iha}. Keeping in mind potential theoretical shortcomings that such an approach might involve we discuss the potential improvement of the LEP+LHC combination in \Sec{sec:WBF}.  We conclude in \Sec{sec:conc}.

\section{Higgs Width Overview In Light Of LHC Results}
\label{sec:hwidth}
Due to its small couplings to light fields, in the SM the Higgs width satisfies $\Gamma_h \ll m_h$ and the narrow width approximation is appropriate for LHC observations of an on-shell Higgs \cite{Passarino:2010qk}.  Specifically, if $\sigma_i$ is the SM prediction for Higgs production in some channel `i' at the LHC and $\text{BR}_j$ is the SM prediction for the branching ratio into a final state `j', then a reasonable approximation for the total cross section in these channels at the LHC is
\begin{eqnarray}
\sigma_{ij} & = &  \frac{c_i^2 c_j^2}{R_h} \sigma_i \text{BR}_j \\
& = & \mu_{ij} \sigma_i \text{BR}_j
\label{eq:lhcobs}
\end{eqnarray}
where we have re-scaled the SM Higgs couplings with some factor which takes the value $c\to1$ in the SM limit, we have similarly rescaled the total decay width by a factor $R_h$, and shown these two may be absorbed into a single `signal-strength' variable $\mu$.\footnote{In reality a simple coupling rescaling is overly simplistic and ideally the effects of new physics above the weak scale should be encoded in higher dimension operators.  However, it is worth noting that the existence of complete models which realize free couplings for the Higgs with SM fields have been demonstrated \cite{Lopez-Val:2013yba}, thus the free-coupling interpretation does have consistent UV-completions.}  We have also assumed that the Higgs width is a free parameter ($\Gamma_h = R_h \Gamma_{SM}$) which is not necessarily given by $\Gamma_h \neq \sum_j c_j^2 \Gamma_j$.  We are free to make this choice because the Higgs may possess additional decay channels into new invisible states or even into hadronic channels which are difficult to detect at the LHC.  This also allows for the fact that couplings to different fields may be altered in uncorrelated ways, for example if the Higgs is coupled to new colored states, e.g. a sequential chiral generation \cite{Kribs:2007nz}, they may significantly modify the $hGG$ coupling at leading order.

\Eq{eq:lhcobs} makes it immediately clear that an unambiguous extraction of the Higgs width is not possible at the LHC from on-shell observations alone as it always appears in combination with Higgs couplings which may also be modified.  Essentially there is a flat direction in parameter space along which observed LHC Higgs signal strengths $\mu_{ij}$ may take the same set of fixed values for a continuous family of width and coupling variations.  For example, taking the form $c_i^2 c_j^2 = R_h$ we have $\mu_{ij}=1$ and an apparently SM-like Higgs even in the presence of modified couplings.  However, this does not imply that no information on the width is obtained. By imposing the assumption of a specific model it is possible to extract constraints.  For example, if it is assumed that the Higgs may not decay to additional invisible particles or to visible particles in new exotic channels, then the total width is given by $\Gamma_h = \sum_j c_j^2 \Gamma_j$ and global fits to the LHC data allow for experimental constraints on the Higgs width \cite{sven,dob,Lopez-Val:2013yba}.  Alternatively, if it is assumed that all couplings are SM-like and the only modification is an increased width due to additional decays then again global fits allow for the extraction of a limit on the Higgs width~\cite{Espinosa:2012vu}.

A complementary approach which relies on combining additional measurements with the LHC on-shell Higgs observations has also been proposed \cite{Caola:2013yja}.  This approach exploits processes in which the Higgs is far off-shell but still plays a role.  In particular in the many parton-level processes contributing to diboson production $pp \to ZZ$ there is one which involves a virtual Higgs: $gg \to h^* \to ZZ$.  This subprocess contributes at a level which is experimentally accessible and hence it is possible to use measurements of $ZZ$ production at high invariant mass to constrain the impact of an off-shell Higgs \cite{Kauer:2013qba,ciaran} (see \cite{ilc} for a related study at a future lepton collider).  This is very useful in the theoretical interpretation of the Higgs properties for a number of reasons \cite{offshella,Offshell}.  The desired application is that if one considers the usual na\"{\i}ve coupling re-scaling of \Eq{eq:lhcobs} then with the Higgs sufficiently off-shell the matrix element does not depend on the Higgs width but simply behaves as $c_{gg} c_{ZZ} \mathcal{M}_{SM}$, where the dependence on the $hGG$ and $hZZ$ couplings is explicit.  Thus experimental measurements of high invariant mass $ZZ$ production can be interpreted as constraints on Higgs couplings.  These constraints can then be combined with measurements of on-shell observables described by \Eq{eq:lhcobs}.  As the off-shell measurement breaks the degeneracy between coupling and width modifications then, under a specific set of assumptions~\cite{offshella}, the combination of on-shell and off-shell measurements allows for an indirect constraint on the total width of the Higgs. Within these limitations both the ATLAS \cite{atlas} and CMS collaborations \cite{Khachatryan:2014iha} have reported limits on the total Higgs width.

There are, however, a number of important caveats and subtleties involved in this mapping from on-shell and off-shell measurements to a width constraint \cite{offshella,ChrisandMichael}. In particular if the Higgs coupling modifications are in any way dependent on the energy at which they are probed the mapping breaks down.  For example, if the $hGG$ coupling is modified by loops of new colored particles with masses of $\mathcal{O}(100\text{'s})$ GeV~\cite{ChrisandMichael}, if new higher dimensional interactions are present~\cite{offshella}, if scalars appear as $s$-channel resonances, or if electroweak symmetry breaking is not SM-like~\cite{offshella,offshellb}, the mapping between the two constraints and thus the interpreted width measurement would be incorrect. While these scenarios can be constrained with other measurements, a generic model-independent interpretation of the width constraint from $gg \to h^* \to ZZ$ is clearly unjustified.  Motivated by these considerations we will now describe a complementary approach which also utilizes processes involving an off-shell Higgs.

Prior to the discovery of the Higgs there were already strong constraints on processes involving an off-shell Higgs from the precision electroweak program at LEP.  Now that the mass of the Higgs is known these LEP constraints may be interpreted as constraints on Higgs couplings, but only under assumptions on the nature of any additional BSM modifications to the precision electroweak observables.  Such LEP Higgs coupling constraints have been considered in a number of works previously \cite{HIGGSEWP,EFTJUST}.  The aspect we will focus on is that the LEP Higgs coupling constraints involve an off-shell Higgs and to a good approximation are not dependent on the Higgs decay width, in analogy with the high invariant mass constraints on $ZZ$ production at the LHC.  Thus the idea is to perform a similar manipulation to the one described previously: Interpret the LEP constraints as Higgs coupling constraints and then feed these back into observables such as \Eq{eq:lhcobs} to extract a bound on the Higgs width.

The constraint is applicable to broad classes of models, however for the sake of demonstrating the use of the LEP+LHC combination we will focus on two specific classes of models as examples, both treating the total Higgs width as a free parameter.  The models are:  
\begin{enumerate}[a)] 
\item Rescaled $hWW$ and $hZZ$ couplings by a factor $c_V$ and with a UV cut-off $\Lambda$.  As in \cite{Baak:2014ora} we will also set the cutoff to $\Lambda = \lambda / \sqrt{|1-c_V^2|}$, motivated by effective theory arguments as described in \cite{EFTJUST}.\footnote{For comparison with the parameterization in \cite{EFTJUST} this parameter is taken to be $\lambda \approx 4 \pi v$.}  Depending on the constraint considered all other couplings may be assumed to be rescaled in the same way, or in some instances they may be taken as free parameters.  We will state which of these two assumptions is taken as and when appropriate. 
\item The Higgs mixed with a singlet scalar in a `Higgs Portal' type of scenario \cite{portal}.  This model introduces two parameters, the mass of the additional scalar $M_S$ and the mixing angle between the SM Higgs and the singlet scalar $\theta$.
\end{enumerate}
As with any indirect constraint the LEP+LHC constraint suffers from its forms of model-dependence, which we will discuss in more detail in the next section.  We consider the two models described above as they are simple examples which are representative of scenarios with modified Higgs couplings, although the applicability of the constraint should go well beyond these models.

\section{LEP as an Off-Shell Higgs Factory}
\label{sec:LEP}
\begin{figure}[t]
\centering
\includegraphics[height=1.35in]{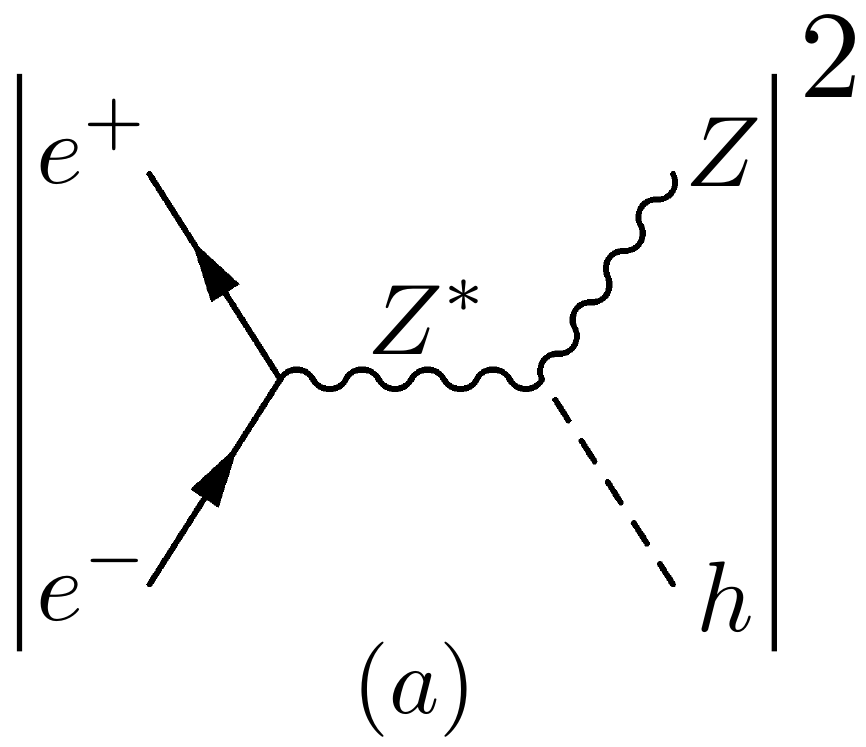} \qquad \qquad \includegraphics[height=1.35in]{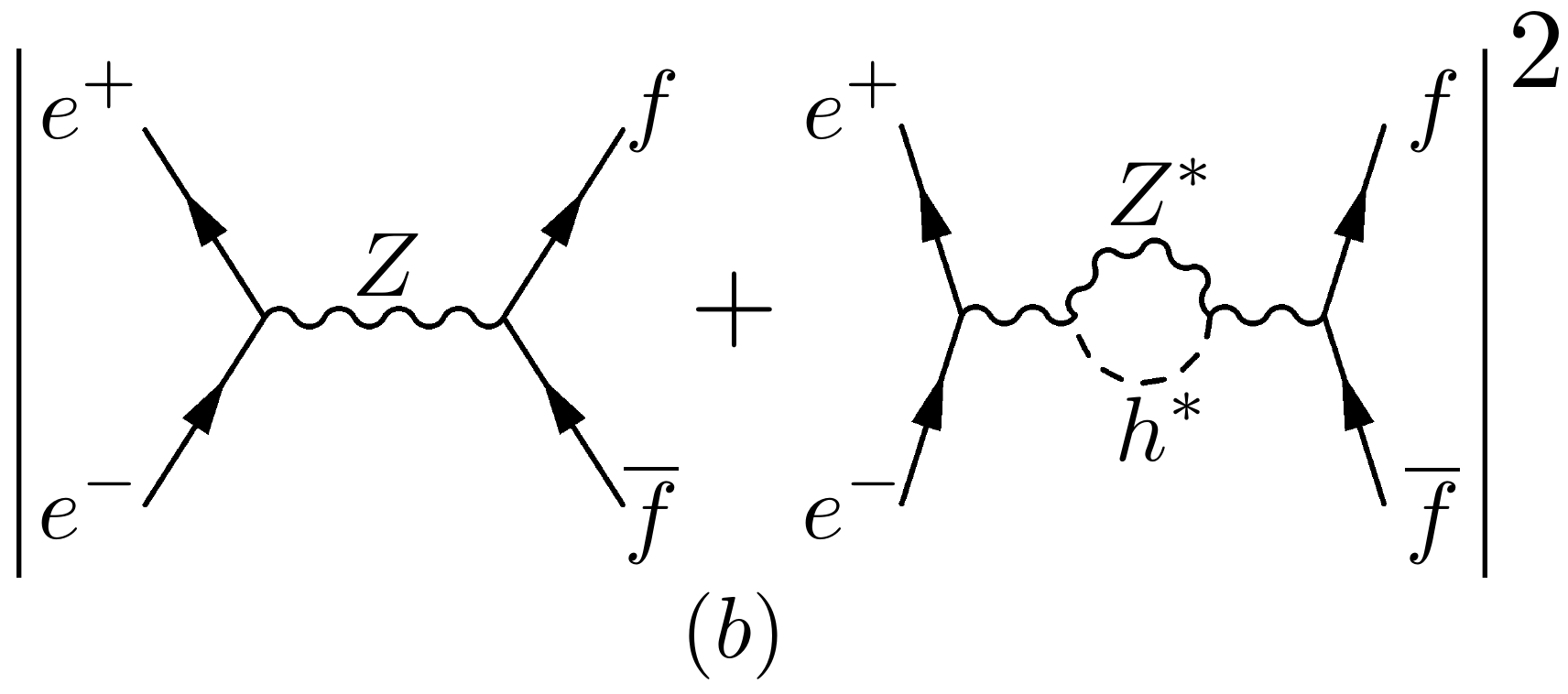}
\caption{(a)  The dominant Higgs production process at a potential future Higgs factory $e^+ e^-$ collider and (b)  $Z$-boson production and decay at an $e^+ e^-$ collider including interference of one loop diagrams involving an off-shell Higgs.  A analogy between the on-shell and off-shell Higgs factory is drawn because the squared amplitude for the Higgs factory is related to the one-loop amplitude which interferes with tree-level diagrams at an off-shell Higgs factory.}
\label{fig:HiggsFactory}
\end{figure}

Currently there is focussed discussion on the possibility of a future `Higgs Factory', an $e^+ e^-$ collider that would produce copious numbers of Higgs bosons in a high precision environment.  The various possibilities under discussion include the ILC \cite{Baer:2013cma}, CLIC \cite{Accomando:2004sz}, FCC-ee (TLEP \cite{Gomez-Ceballos:2013zzn}), and CEPC \cite{cepc}.  The common feature among these colliders is that in the initial lower energy stages the dominant production mechanism for the Higgs is associated production $e^+ e^- \to h Z$.  This is depicted in \Fig{fig:HiggsFactory} (a).  In terms of diagrams, by squaring \Fig{fig:HiggsFactory} (a) we arrive at the interference term of \Fig{fig:HiggsFactory} (b).  Thus the one-loop corrections to LEP observables are a close cousin to the on-shell production at a Higgs factory.  By this connection one could consider LEP as an `off-shell' Higgs factory and the constraints from the LEP measurements can be interpreted as precision constraints on off-shell Higgs processes.

The Peskin-Takeuchi parameters \cite{Peskin:1990zt} cleanly frame the LEP constraints on modifications of the SM electroweak sector.  Thus the off-shell Higgs constraints are best presented in terms of the $S-T-U$ parameters.  We assume a common modification to the $hZZ$ and $hWW$ couplings.  This choice is not exhaustive of new physics at or above the weak scale as a number of higher dimension operators may enter at tree-level and one-loop as will be discussed in more detail at the end of this section.  However this choice is appropriate for the specific models considered here.  The contributions of a Higgs-like scalar of mass $m_H$, with couplings to vector bosons $g = c_V \times g_{\text{SM}}$, to the $S-T-U$ parameters are
\be
s_H (m_H,c_V)  & = & \frac{c_V^2}{\pi M_Z^2} \bigg( B^{OO}_{M_Z^2,Z}- B^{OO}_{0,Z}  - M_Z^2 \big(B^{O}_{M_Z^2,Z} - B^{O}_{0,Z} \big) \bigg)  \\
t_H (m_H,c_V)  & = & \frac{c_V^2}{4 \pi M_W^2 S_W^2} \bigg( B^{OO}_{0,W}- B^{OO}_{0,Z} + M_Z^2 B^{O}_{0,Z} - M_W^2 B^{O}_{0,W} \bigg)  \\
u_H (m_H,c_V) &  =  & \frac{c_V^2}{\pi} \bigg( \big(B^{O}_{M_Z^2,Z} - B^{O}_{0,Z} \big) - \big(B^{O}_{M_W^2,W} - B^{O}_{0,W} \big)-\frac{1}{M_Z^2} \big( B^{OO}_{M_Z^2,Z}- B^{OO}_{0,Z} \big)  \nonumber \\
& &  +\frac{1}{M_W^2} \big( B^{OO}_{M_W^2,W}- B^{OO}_{0,W} \big) \bigg)
\label{eq:fulla}
\ee
where the full loop functions are
\be
B^{O}_{0,V} & = & 2 \frac{\kappa^2 \log \kappa-(1-\kappa^2) \log \gamma }{1-\kappa^2}\nonumber \\
B^{OO}_{0,V} & = & M_H^2 \frac{\kappa^4 \log \kappa^4+(1-\kappa^4)  \left(1-4 \log \gamma \right)}{8 (1-\kappa^2)} \nonumber\\
B^{O}_{M_V^2,V} & = & \frac{1}{2 \kappa^2} \Bigg( (1-2 \kappa^2) \log \kappa^2 + 2 \kappa^2 (1-2 \log \gamma) +  \sqrt{1-4 \kappa^2} \log \left(\frac{1-2 \kappa^2 + \sqrt{1-4 \kappa^2}}{2 \kappa^2} \right) \Bigg) \nonumber \\
B^{OO}_{M_V^2,V} & = & \frac{M_H^2}{36 \kappa^4} \Bigg( \kappa^2 (4 \kappa^4 + 18 \kappa^2-3)  + \frac{3}{2} (1-4 \kappa^2)^{3/2} \log \left(\frac{1-2 \kappa^2 - \sqrt{1-4 \kappa^2}}{2 \kappa^2} \right)  \nonumber \\
& & - 3 (1-6 \kappa^2 + 6 \kappa^4 + 4 \kappa^6) \log \kappa  - 6 \kappa^4 (3+2 \kappa^2) \log \gamma \Bigg) .
\label{eq:loop}
\ee
$\kappa=M_V/M_H$ and $\gamma = M_H/\Lambda$, where $\Lambda$ is the $\overline{\text{MS}}$ UV cutoff.

\begin{figure}[t]
\centering
\includegraphics[height=2.6in]{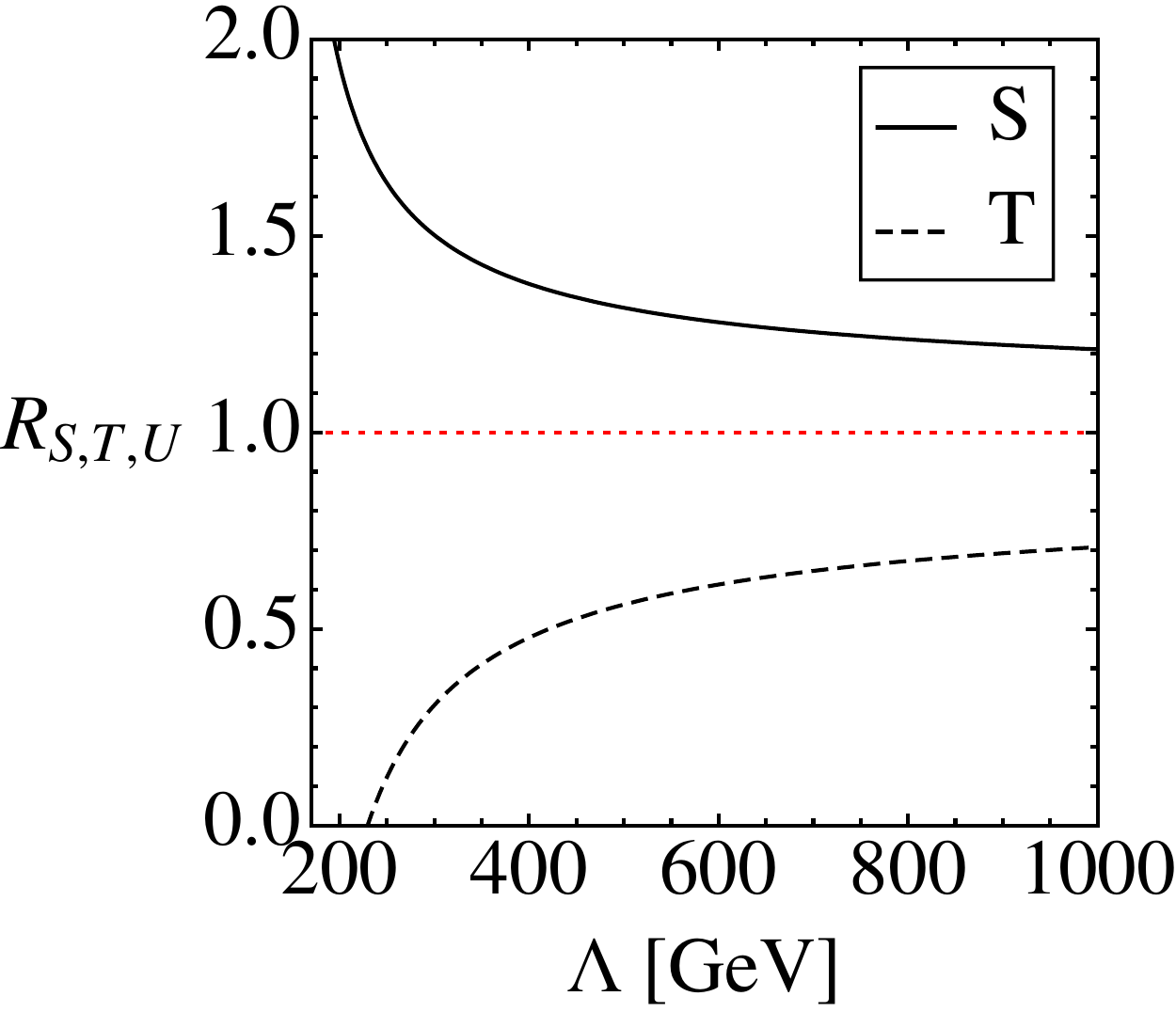}
\caption{The ratio of full loop functions in \Eq{eq:full} with the leading-log expressions in \Eq{eq:LL} as a function of the UV cutoff $\Lambda$.  It is clear that the leading-log expressions may underestimate ($S$) or overestimate ($T$) the corrections by more than $20\%$, even with a cutoff extending to $1$ TeV where one might expect the leading-log expression to be accurate.  However, the leading-log result is often adequate for estimates as it is the dominant contribution for $\Lambda \gtrsim 300$ GeV.}
\label{fig:effrat}
\end{figure}
 
For the model with re-scaled Higgs couplings the deviations of the $S$, $T$, and $U$ parameters from the Standard Model prediction are
\be
S & = & s_H (m_h,c_V)-s_H (m_h,1) \nonumber \\
T & = & t_H (m_h,c_V)-t_H (m_h,1)  \nonumber \\
U & = & u_H (m_h,c_V)-u_H (m_h,1).
\label{eq:full}
\ee
In the limit with modified couplings and a large UV cutoff the leading-log (LL) approximation to the electroweak precision parameters may be found directly from the $\log \gamma^2$ terms of \Eq{eq:loop}.  These are
\be
S_{\text{LL}} & = &  -\frac{1}{6 \pi} (1-c_V^2) \log \gamma  \nonumber \\
T_{\text{LL}} & = & \frac{3 M_Z^2}{8 \pi M_W^2} (1-c_V^2) \log \gamma  \nonumber \\
U_{\text{LL}} & = &  0
\label{eq:LL}
\ee
in agreement with \cite{Baak:2014ora}.  It is interesting to consider the full expression versus the leading-log approximation.  In \Fig{fig:effrat} we show the S and T corrections relative to the leading-log approximation as a function of the cutoff $\Lambda$.  It is clear that for most purposes the leading-log approximation is adequate, however the full loop expressions are desirable for accurate results.

\begin{figure}[t]
\centering
\includegraphics[height=2.6in]{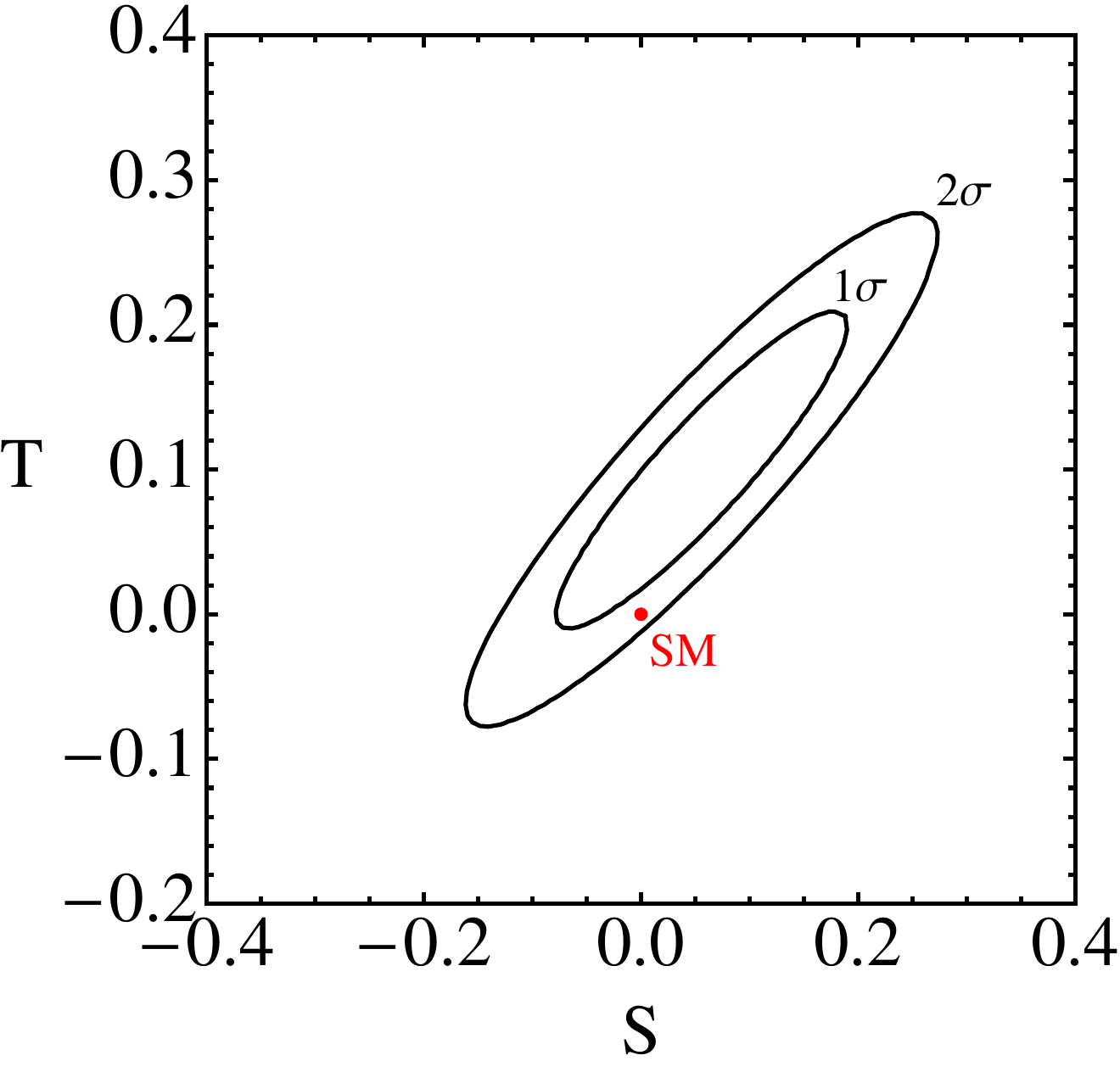}
\caption{The standard $S-T$ ellipse for $U=0$ with the SM point depicted at $S=T=0$.}
\label{fig:STU}
\end{figure}

In the Higgs portal model the $S$-$T$-$U$ expressions are
\be
S & = & s_H (m_h,\cos{\theta})+s_H (M_S,\sin{\theta})-s_H (m_h,1)\\
T & = & t_H (m_h,\cos{\theta})+t_H (M_S,\sin{\theta})-t_H (m_h,1) \\
U & = & u_H (m_h,\cos{\theta})+u_H (M_S,\sin{\theta})-u_H (m_h,1) 
\ee
In both of the models there is an additional parameter (either $\Lambda$, or $M_S$) in addition to the modified Higgs coupling which must be considered.  This is essentially due to the fact that the LEP constraints are logarithmically sensitive to UV physics.  This is an additional element of model-dependence unique to these coupling constraints which the off-shell $gg\to h^* \to ZZ$ constraints do not suffer from.

\begin{figure}[t]
\centering
\includegraphics[height=2.7in]{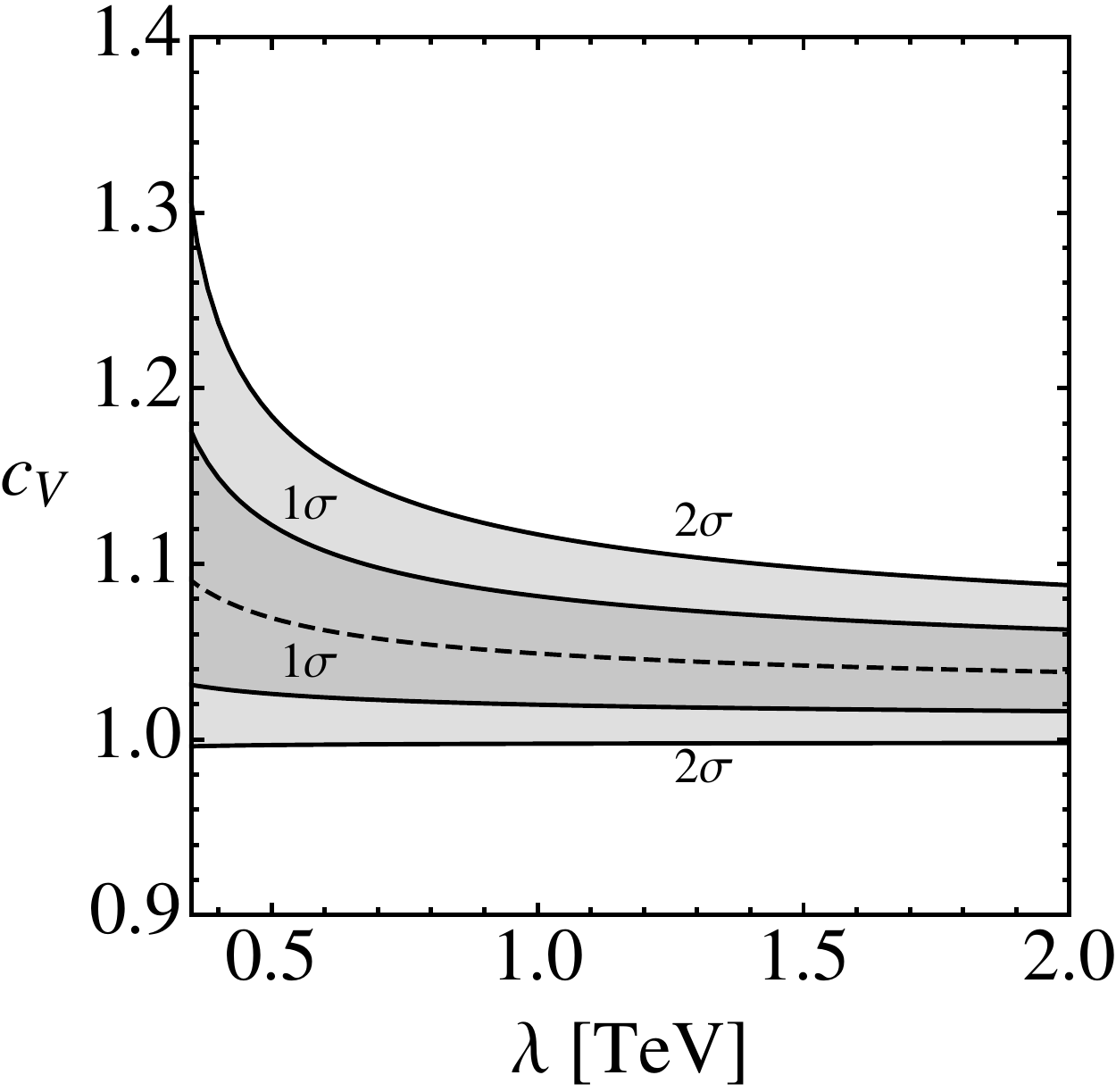} \qquad \includegraphics[height=2.7in]{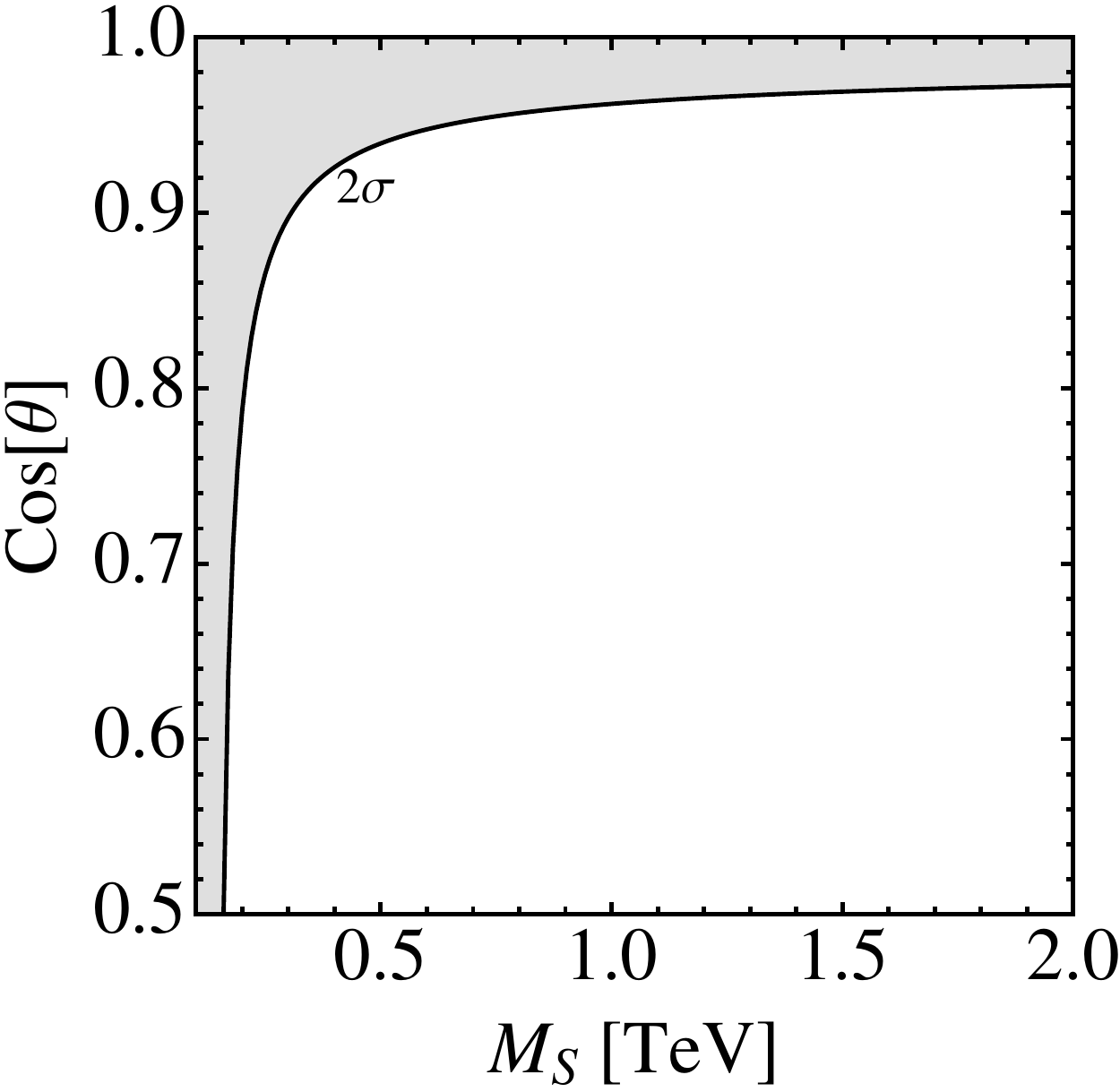}
\caption{LEP constraints on modified Higgs couplings in the two models described in \Sec{sec:hwidth}.  There is already tension $1 \sigma$ between precision electroweak fits and the SM, hence in the models model with re-scaled couplings (left panel) the SM limit is only within the $2 \sigma$ contours.  For the Higgs portal model (right panel) we only show the $2 \sigma$ contour as the $1 \sigma$ contour is very close.}
\label{fig:coupSTU}
\end{figure}

We use the central values, errors, and correlation matrix for the $S$-$T$-$U$ parameters from \cite{Baak:2014ora}.  Using these constraints we find the regions of parameter space allowed by LEP data.  In \Fig{fig:STU} we show the standard $S-T$ ellipse under the assumption that $U=0$.  This agrees well with similar figures in \cite{Baak:2014ora}.

In \Fig{fig:coupSTU} we plot the constraints on both of the models from LEP measurements.  This figure makes it clear that the LEP measurements are effective in constraining modified Higgs couplings.  This point has been emphasized previously by many authors \cite{HIGGSEWP,EFTJUST}.  The most relevant point for this work which has not been emphasized previously is that these coupling constraints are valid irrespective of the Higgs decay width, thus they can later be combined with LHC on-shell Higgs observations in order to determine indirect constraints on the decay width.

Before proceeding it is worthwhile pausing to consider scenarios in which these constraints are valid.  Essentially the underlying assumption is that only the Higgs sector of the SM has been modified.  In practice this requires that there are no BSM modifications of weak gauge-boson self-energies or interactions.  It also requires that modifications of fermion couplings to weak gauge bosons are also absent.  Thus in the EFT language all operators which do not involve the Higgs doublet are absent.  Furthermore, it assumes that the Lorentz structure of the modified Higgs-gauge boson couplings is precisely the same as for the SM tree-level couplings.  This restricts to models of the type where the Higgs boson is mixed with e.g. another singlet scalar, or in EFT language restricts to higher dimension operators which modify couplings universally, such as $(\partial |H|^2)^2$.  These are simply the restrictions which apply to the commonly adopted `$\kappa$-framework' in which Higgs couplings are rescaled by some overall constant factor, thus the theoretical underpinning of this work is on a similar footing to the application of this framework for analyses of Higgs couplings in off-shell or boosted regimes.  However, we would like to emphasize that as with any indirect constraint on new physics, the above-mentioned assumptions should be kept in mind throughout and they may break down in some models.

\section{Combining LEP measurement with LHC8 Data}
\label{sec:currentcombo}
In \Sec{sec:LEP} it was demonstrated that constraints on virtual off-shell Higgs corrections at LEP lead to constraints on Higgs couplings which do not depend on the width.  On-shell Higgs measurements at the LHC have already placed strong constraints on the overall signal strength $\mu$.  In this section we combine the two to determine current constraints on the total Higgs decay width.

The off-shell LEP constraints are of the form
\be
c_{min} < c_V < c_{max},
\ee
where $c_{min}$ and $c_{max}$ may depend on model parameters.  The on-shell LHC constraints are of the form
\be
\mu_{min} < \mu \left(=\frac{c_V^4}{R_h} \right) < \mu_{max}.
\ee
Thus we may rearrange these inequalities to determine indirect constraints on the Higgs width\footnote{Similar inequalities have been discussed in \cite{dob}.}
\be
\frac{c_{min}^4}{ \mu_{max}} < R_h < \frac{c_{max}^4}{\mu_{min}} ~~.
\label{eq:comb}
\ee
Ideally the LEP and LHC constraints would be merged into a combined likelihood function to allow for a more sophisticated statistical analysis, however the simple combination of \Eq{eq:comb} serves the purpose of illustrating the use of LEP measurements to determine a constraint on the Higgs width, and in any case the quantitative results should be approximately representative of the precision achievable.

\begin{figure}[t]
\centering
\includegraphics[height=2.6in]{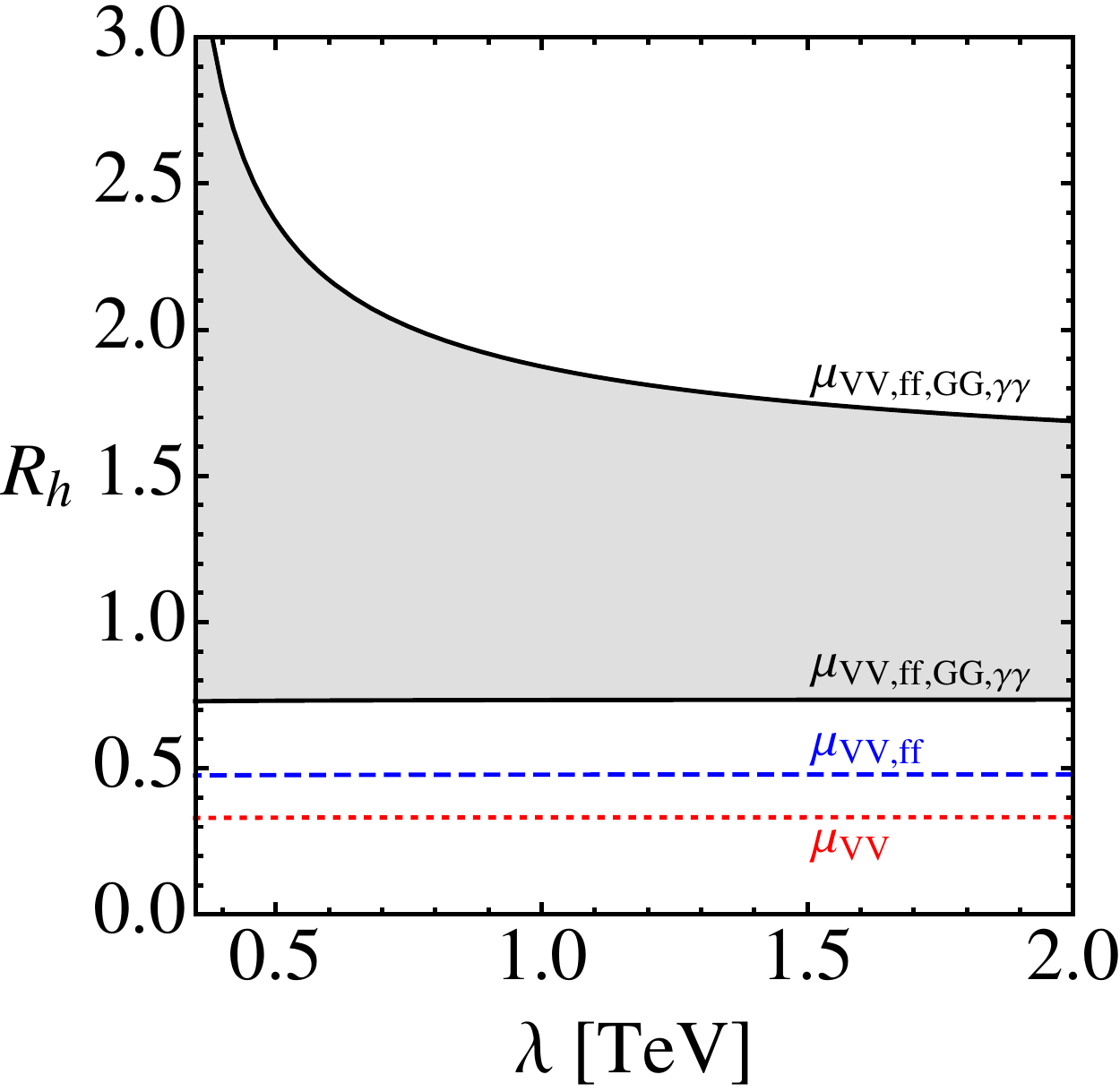} \qquad \includegraphics[height=2.6in]{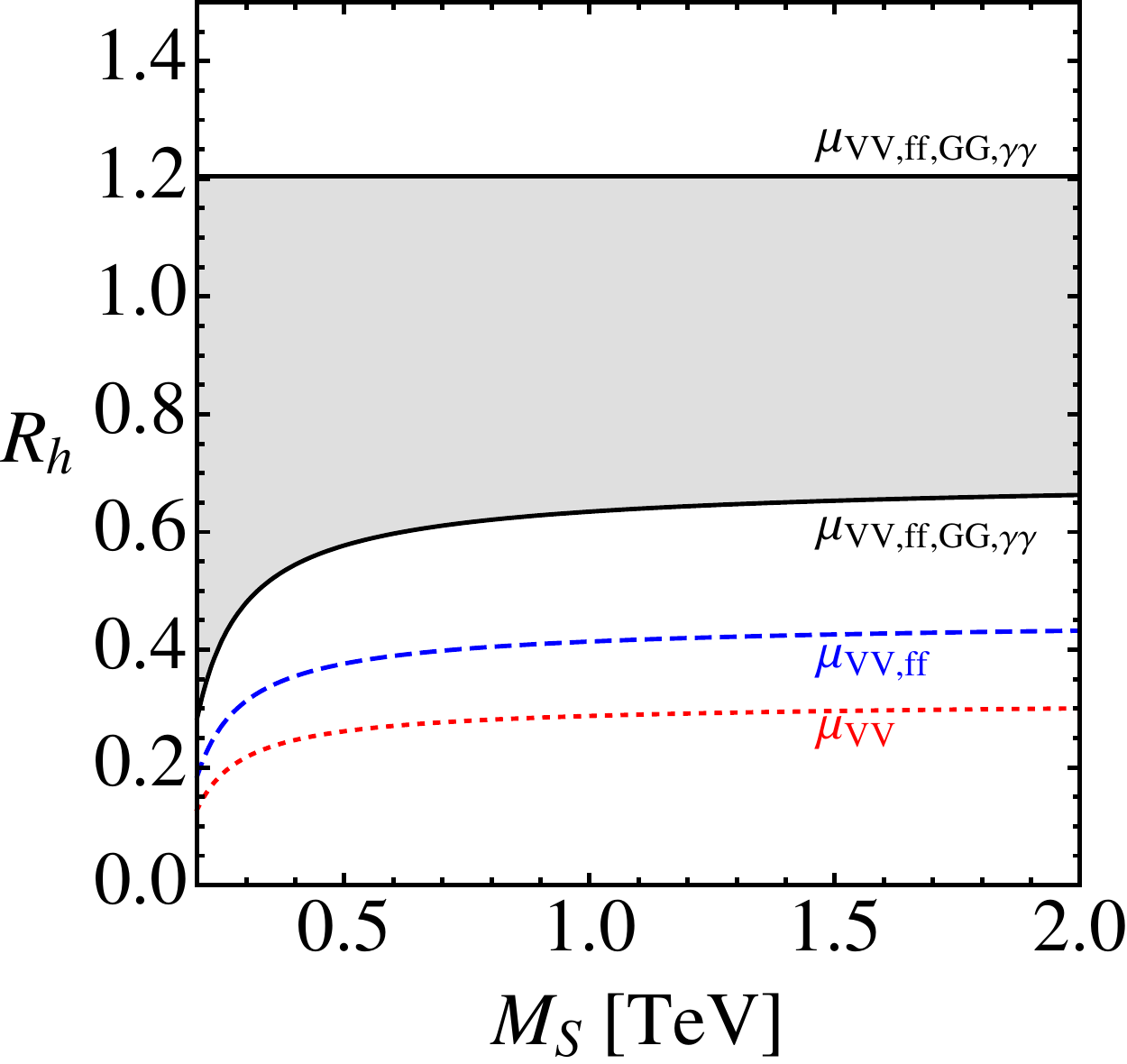}
\caption{Combined $2 \sigma$ LEP and current LHC signal strength constraints on the Higgs width in the model with modified couplings and a UV cutoff (left) and a model with a mixed-in singlet scalar (right).  Please see the surrounding text for an explanation of the various contours.}
\label{fig:widthconstLHC}
\end{figure}

\subsection{Current Limits: Universal Coupling Rescaling}
\label{sec:univ}
Combining all of the observed channels the current status of the mean and uncertainties in the overall Higgs signal strength are: ATLAS ($\mu = 1.18^{+0.15}_{-0.15}$) and CMS ($\mu = 1.00^{+0.14}_{-0.14}$) (see e.g.\ \cite{Duehrssen}).  We make an unofficial signal strength combination of $\mu \approx 1.09^{+0.13}_{-0.13}$, where we have combined the statistical errors as standard in quadrature, improving this source of error by a factor $\sim 1/\sqrt{2}$, however the error has not improved significantly as we have assumed systematic and theoretical errors do not improve upon the combination of results from both experiments.  Thus from this approximate combination at $2 \sigma$ confidence level we take $\mu_{min} \approx 0.83$, $\mu_{max} \approx 1.35$.  Using \Eq{eq:comb} we may combine these limits with the LEP constraints on the Higgs couplings shown in \Fig{fig:coupSTU} to find the current constraints on the Higgs width in these two models.

In \Fig{fig:widthconstLHC} we show the $2 \sigma$ confidence contours on the total Higgs for both of the example models described in \Sec{sec:hwidth}.  The constraint from current LHC on-shell signal strength constraints is labelled $\mu_{VV,ff,GG,\gamma\gamma}$.  For reference the $2 \sigma$ limits at new physics scales of $1$ TeV for a model with a universal rescaling of couplings are $0.73 \lesssim R_h (\lambda=1 \text{ TeV}) \lesssim 1.87$, which is already competitive with constraints using other methods \cite{atlas,Khachatryan:2014iha}.  For the model of a mixed-in singlet scalar  the limits are $0.63 \lesssim R_h (M_S=1 \text{ TeV}) \lesssim 1.20$.  It is worth noting that in this case the upper limit on the Higgs width is only strong due to the theoretical limitation $\cos (\theta) \leq 1$, thus in \Eq{eq:comb} we have $R_h < 1/\mu_{min}$.

\subsection{Current Limits: $hGG$- and $h\gamma \gamma$-Independent Combination}
\label{sec:hgg}
It is possible to construct less model-dependent constraints by focussing on specific channels.  For example, we may constrain the Higgs width in a model in which the $hVV$ and $h\overline{f}f$ couplings have been modified by the same factor and the $hGG$ and $h\gamma\gamma$ couplings are allowed to be rescaled independently as free parameters, thus this is applicable to scenarios where the Higgs may be coupled to new colored fields.\footnote{An example of such a scenario would be the Higgs portal mixing with a singlet and Higgs couplings to new colored scalars.}  This is achieved by choosing production channels and final state decays that are independent of the $hGG$ and $h\gamma\gamma$ couplings.  We consider the LHC constraints from Higgs associated production and $\overline{b}b$ decays, $hV,h\to \overline{b}b$.  The best fit signal strength in this channel is $\mu = 1.01^{+0.53}_{-0.5}$ \cite{Khachatryan:2014jba}.\footnote{It should be kept in mind that the current constraints from WBF production with subsequent Higgs decay to taus are marginally stronger than for associated production.  However in this channel for typical cuts there is a significant contribution from gluon fusion.  In \Sec{sec:WBF} we demonstrate that it is possible in future to almost fully remove this gluon fusion contribution, enabling $hGG$-independent constraints to be set with the production channel also.}

In \Fig{fig:widthconstLHC} the $2 \sigma$ confidence contour for this constraint is labelled $\mu_{VV,ff}$.  An upper bound is not realized as the signal strength in $hV,h\to \overline{b}b$ is consistent with zero at $2 \sigma$ confidence, thus by \Eq{eq:comb} the width is consistent with very large values.  For reference the $2 \sigma$ width lower limit at new physics scales of $1$ TeV for a model with any value of $hGG$ and $h\gamma \gamma$ couplings is $0.48 \lesssim R_h (\lambda=1 \text{ TeV})$.  For the model of a mixed-in singlet scalar the lower limit is $0.41 \lesssim R_h (M_S=1 \text{ TeV})$.  These limits are less model-dependent than the limits from combined global signal strength constraints, making them applicable to a larger range of models.  However this has come at a price as the more model-independent limits are quantitatively weaker.

\subsection{Current Limits: $hGG$-, $h\gamma \gamma$-, and $h\overline{f} f$-Independent Combination}
\label{sec:hff}
We may extend the model-independence even further by combining LEP constraints on the $hVV$ couplings with current LHC constraints on production and decay channels which only feature the $hVV$ couplings.  This essentially leads to a Higgs width constraint in the `$\kappa$-framework' language which is independent of all couplings.  To this end we again choose associated production followed by Higgs decays to $W$-bosons, $hV,h\to W^+ W^-$.  The best fit signal strength in this channel is $\mu = 0.80^{+1.09}_{-0.93}$ \cite{Khachatryan:2014jba}.

In \Fig{fig:widthconstLHC} the $2 \sigma$ confidence contour for this Higgs width bound is labelled $\mu_{VV}$.  Again there is no upper bound.  For reference the $2 \sigma$ width lower limit at new physics scales of $1$ TeV for a model with any value for the $hGG$, $h\gamma \gamma$, and $h\overline{f} f$ couplings is $0.33 \lesssim R_h (\lambda=1 \text{ TeV})$.  For the model of a mixed-in singlet scalar the lower limit is $0.29 \lesssim R_h (M_S=1 \text{ TeV})$.  These limits are even less model-dependent than the limits in the previous two sections.  In fact, other than the well-motivated assumption that the $hWW$ and $hZZ$ couplings are scaled in the same way, these limits allow for all Higgs couplings to scale freely and independently and thus have a very reduced model dependence.  However the theoretical limitations discussed at the end of \Sec{sec:LEP} still apply.

\subsection{Higgs width constraints in other models}
\label{sec:other}
Finally we note that in general models with modified Higgs couplings the possible coupling variations may be very rich.  However in a specific model, such as a Two Higgs Doublet Model, it would be possible to perform exactly the same procedure to determine constraints on the Higgs width.  For a given parameter choice the LEP constraints on the model, including one-loop contributions from any Higgs-like states, may be combined with observed Higgs signal strengths at the LHC to determine limits on the Higgs width.

\section{Looking to the future:  Vector-only LHC Higgs Constraints}
\label{sec:WBF}
As discussed in \Sec{sec:currentcombo} we would ideally like to construct a constraint which is as independent of as many couplings as possible to reduce the model-dependence of the width extraction.  To this end we use a projection of vector boson-only production and decay mode approaches~\cite{ChrisandMichael} at the LHC (see also \cite{Kauer:2013qba,vbfellis}).  Specifically, we analyze the processes $pp \to (h \to W^+ W^-\to l^+l^- \nu \bar{\nu}) jj$, $pp \to (h \to ZZ^*\to l^+l^- \nu \bar{\nu}) jj$ and $pp \to (h \to ZZ^*\to 4l) jj$.  This choice is advantageous as these LHC measurements constrain the same $hVV$ Higgs couplings as the LEP constraint, thus when these observations are combined this will result in a much more robust constraint on the Higgs width in the context of the models considered.  In particular any dependence on the $hGG$ coupling is essentially removed, implying that the final width constraint will apply to any model with a modified $hVV$ coupling and a modified width.\footnote{Again, subject to the limitations discussed in \Sec{sec:LEP}.}  Let us first construct the LHC on-shell observables.

We use {\sc{Vbfnlo}} v2.7~\cite{vbfnlo} to simulate the weak boson fusion and gluon fusion events for full leptonic final states at 14 TeV. We use a leading order RGE-improved mode of {\sc{Vbfnlo}} that uses the $t$-channel momentum transfer as the relevant scale for parton distributions and strong coupling running \cite{wbfscales}, and pre-select the Higgs on-shell region.  

Subsequently the {\sc{Vbfnlo}} events are showered and hadronised with {\sc{Herwig++}} \cite{Bahr:2008pv}. For the backgrounds we consider continuum $ZZ$, $WW$ and $WZ$ production including all interference effects, generated with {\sc{MadGraph}} \cite{Alwall:2011uj}, as well as $t\bar t$ production generated using {\sc{Alpgen}}~\cite{alpgen}. Detector effects and reconstruction efficiencies are included and based on the ATLAS Krakow parametrization~\cite{detector}. 

Jets are reconstructed with the anti-$k_T$ jet clustering algorithm~\cite{Cacciari:2008gp} with $p_T>35$ GeV, $|y_j|<5.0$ and resolution parameter $R=0.4$. We again adopt the ATLAS Krakow parametrization~\cite{detector} to include jet resolution effects, $b$-jet efficiencies and fake rates. We consider light charged leptons (i.e. electrons and muons) to be isolated if $p_{T,l} > 15$ GeV, $|y_l| < 2.5$, and if the hadronic energy deposit within a cone of size $R = 0.3$ is smaller than $10\%$ of the lepton transverse momentum.

In the Higgs portal model there may be additional contributions to the Higgs signal region from the heavy scalar.  We have thus included the heavy scalar in the simulation and checked that the signal contamination is negligible for the masses and mixing angles that satisfy the LEP constraints, thus it is self-consistent to only consider the 125 GeV Higgs-like scalar contributions in this section.

\subsection{$hjj \to WW^*jj \to l^+ l^- \nu \bar{\nu}jj$}
\label{sec:hww}

For the $h\to WW\to 2l+\slashed{E}_T$ decay mode of WBF production we require exactly two isolated leptons with $\Delta R_{j,l} > 0.4$ and reject events with $80 < m_{l_1l_2} <100$ GeV to discriminate from $h\to ZZ$. We impose the following WBF cuts on the two hardest jets
\begin{alignat}{5}
\label{eqn:vbf}
y_{j_1} \times y_{j_2} < 0, \quad |y_{j_1} - y_{j_2}| > 4.5, \quad m_{j_1j_2} > 800~\text{GeV}.
\end{alignat}
To further suppress the backgrounds we force the Higgs to be central by requesting~\cite{Kauer:2000hi}
\begin{eqnarray}
\label{eqn:raplep}
\min(y_{j_1},y_{j_2}) < y_{l_1},y_{l_2} < \max(y_{j_1},y_{j_2}).
\end{eqnarray}
At this point the dominant background is $t\bar{t}$, see Tab.~\ref{tab:ww}. To further improve $S/B$ we veto events with a $b$-tagged jet. To achieve $S/B \sim 1/3$ we require that there be no additional jet between the two tagging jets, i.e. $\min(y_{j_1},y_{j_2}) < y_j < \max(y_{j_1},y_{j_2})$~\cite{Barger:1994zq}. Finally, we isolate the Higgs peak via the transverse mass
\begin{equation}
\label{eq:mT2l}
  m_{T,2l}^2 = \left [ \sqrt{m^2_{l_1l_2} + p^2_{T,ll} } + |p_{T,\text{miss}}| \right ]^2 -
   \left [ {\vec{p}}_{T,ll} + {\vec{p}}_{T,\text{miss}} \right ]^2,
\end{equation}
by requiring $80 \leq m_{T,2l} \leq 150$ GeV and obtain $S/B \sim 1$.

\begin{table*}[!t]
\label{tab:ww}
\parbox{1.0\textwidth}
{\begin{center}
\begin{tabular}{|c|c|c|c|c|c|}
\hline
 Sample & lepton cuts & WBF cuts & $b$-veto & jet veto & $m_{T,2l}$ cut \\
\hline\hline
 $(h\to WW) jj$~WBF &  2.803    & 1.015      &   0.996   &    0.958   & 0.561 \\
 $(h\to WW) jj$~GF &  0.887   & 0.105            &   0.101   &    0.069  & 0.039   \\
 $t\bar{t}+$jets &   18189.60                &      24.779        &   6.496       &       0.910  & 0.279   \\
 $WW/WZ/ZZ+$jets &  556.545               &     3.019     &    2.818           & 1.635    & 0.344  \\
 \hline
\end{tabular}
\end{center}}
\hskip 0.6cm
\parbox{1.0\textwidth}{
    \caption{Results for 2 leptons + $\slashed{E}_T$ search. The cross sections are given
      in femtobarns, corresponding to proton-proton collisions at
      $\sqrt{s} = 14$ TeV. Further details on the cuts can be found in
      the text of Sec.~\ref{sec:hww}.}
  }
\end{table*}

\subsection{$hjj \to ZZ^*jj \to l^+ l^- \nu \bar{\nu}jj$}
\label{sec:zz2l}
For the $h\to ZZ\to 2l+\slashed{E}_T$ final state in WBF, we again require exactly 2 isolated leptons.  In this case we impose $85 < m_{l_1l_2} < 95$ GeV to isolate the $Z\to l^+l^-$ decay. After that we proceed with imposing the WBF cuts of Eqs.~(\ref{eqn:vbf})-(\ref{eqn:raplep}).

As in Sec.~\ref{sec:hww} we further improve $S/B$ by imposing a $b$-jet veto and we reject events if there is an additional jet between the two tagging jets. With $80 \leq m_{T,2l} \leq 150$ GeV, as defined in Eq.~\ref{eq:mT2l}, we find $S/B\sim 1/3$ for events that pass all cuts, see Table~\ref{tab:zz2l}.

\begin{table*}[!t]
\label{tab:zz2l}
\parbox{1.0\textwidth}
{\begin{center}
\begin{tabular}{|c|c|c|c|c|c|}
\hline
 Sample & lepton cuts & WBF cuts & $b$-veto & jet veto  & $m_{T,2l}$ cut \\
\hline\hline
 $(h\to ZZ^* \to l^+l^- \nu \bar{\nu}) jj$~WBF &  0.151    & 0.061     &   0.060   &    0.057  & 0.035    \\
  $(h\to WW^* \to l^+l^- \nu \bar{\nu}) jj$~WBF & 0.065    & 0.029     &   0.028   &    0.026 &0.016 \\
 $(h\to ZZ^* \to l^+l^- \nu \bar{\nu}) jj$~GF &  0.044   & 0.005            &   0.004  &    0.003    & 0.002 \\
 $t\bar{t}+$jets &   1667.33               &     2.051        &   0.539      &      0.073   & 0.025  \\
 $ZZ/WZ/WW+$jets &  81.822                &     0.319     &    0.310          & 0.168    & 0.075 \\
 \hline
\end{tabular}
\end{center}}
\hskip 0.6cm
\parbox{1.0\textwidth}{
    \caption{Results for $h \to ZZ^*$ in the 2-lepton + $\slashed{E}_T$ final state.  Further details on the cuts can be found in
      the text of Sec.~\ref{sec:zz2l}.}
  }
\end{table*}

\subsection{$hjj \to ZZ^*jj \to l^+ l^- l'^+ l'^-jj$}
\label{sec:4l}
In the $h\to ZZ\to \text{charged leptons}$ channel we require exactly 4 isolated leptons with $\Delta R_{j,l} > 0.4$ and impose slightly weaker WBF cuts compared to Secs.~\ref{sec:hww} and \ref{sec:zz2l} to retain more signal
\begin{eqnarray}
\label{eqn:vbfzz}
y_{j_1} \times y_{j_2} < 0, \quad |y_{j_1} - y_{j_2}| > 4.5, \quad m_{j_1j_2} > 600~\text{GeV}.
\end{eqnarray}
After $b$- and jet vetos, as outlined in Secs.~\ref{sec:hww} and \ref{sec:zz2l}, the invariant mass of the 4 leptons has to be in a window $115 \leq m_{4l} \leq 135$ GeV. At the expense of a low signal yield we find $S/B \gg 1$.

\begin{table*}[!t]
\label{tab:zz4l}
\parbox{1.0\textwidth}
{\begin{center}
\begin{tabular}{|c|c|c|c|c|c|}
\hline
 Sample & lepton cuts & WBF cuts & $b$-veto & jet veto  & $m_{4l}$ cut \\
\hline\hline
 $(h\to ZZ^* \to 4l) jj$~WBF & 0.012  & 0.006     &   0.005   &    0.005  & 0.005     \\
 $ZZ/WZ/WW+$jets &  4.688                 &     0.031    &    0.030          &0.020    & $<0.001$ \\
 \hline
\end{tabular}
\end{center}}
\hskip 0.6cm
\parbox{1.0\textwidth}{
    \caption{Results for $h \to ZZ^*$ in the 4-lepton final state.  Further details on the cuts can be found in
      the text of Sec.~\ref{sec:4l}.}
  }
\end{table*}

\subsection{Combining Channels}
To determine the possible future LHC Higgs signal-strength constraints we use the above signal and background cross sections to determine statistical uncertainties achievable.  We treat the small gluon fusion contribution as background.  In a particular channel `i' the total statistical Higgs signal strength uncertainty is taken as
\be
\Delta \mu_i = \frac{\sqrt{\mathcal{L} (\sigma_{BG,i}+\sigma_{i})}}{\mathcal{L} \sigma_{i}} ~~,
\ee
where $\mathcal{L}$ is the integrated luminosity, $\sigma_{BG,i}$ is the background cross section and $\sigma_{i}$ is the SM Higgs signal cross section.  We estimate the combined statistical uncertainty as
\be
\Delta \mu = \frac{1}{\sqrt{\sum_i 1/(\Delta \mu_i)^2}} ~~.
\ee
Combing all of the channels leads to an expected signal-strength statistical uncertainty of $\Delta \mu = 10 \% ~ (3 \%)$ with an integrated luminosity of $\mathcal{L} = 300~\text{fb}^{-1}$ ($3~\text{ab}^{-1}$).  These numbers are based solely on statistical uncertainties and are thus not conservative.  There are many sources of systematic uncertainty and the potential leading source is likely to come from jet vetoes.  We do not have accurate estimates of the systematic uncertainty thus we will take three benchmark scenarios motivated by the statistical uncertainties described above.  These benchmarks are $\Delta \mu = 3\%,  10\%, 20\%$.  The first estimate is a maximally optimistic estimate which assumes zero systematic error at the $3~\text{ab}^{-1}$ HL-LHC, the second is likely to be more realistic for the $3~\text{ab}^{-1}$ HL-LHC if systematic errors were reduced to the $\sim 10\%$ level, this benchmark is also motivated by the statistics-only scenario for the $300~\text{fb}^{-1}$ LHC, and in the third we have doubled this uncertainty to demonstrate the impact systematic uncertainties may have.

\subsection{Reducing the Model Dependence in Future Higgs Width Constraints}
\begin{figure}[t]
\centering
\includegraphics[height=2.6in]{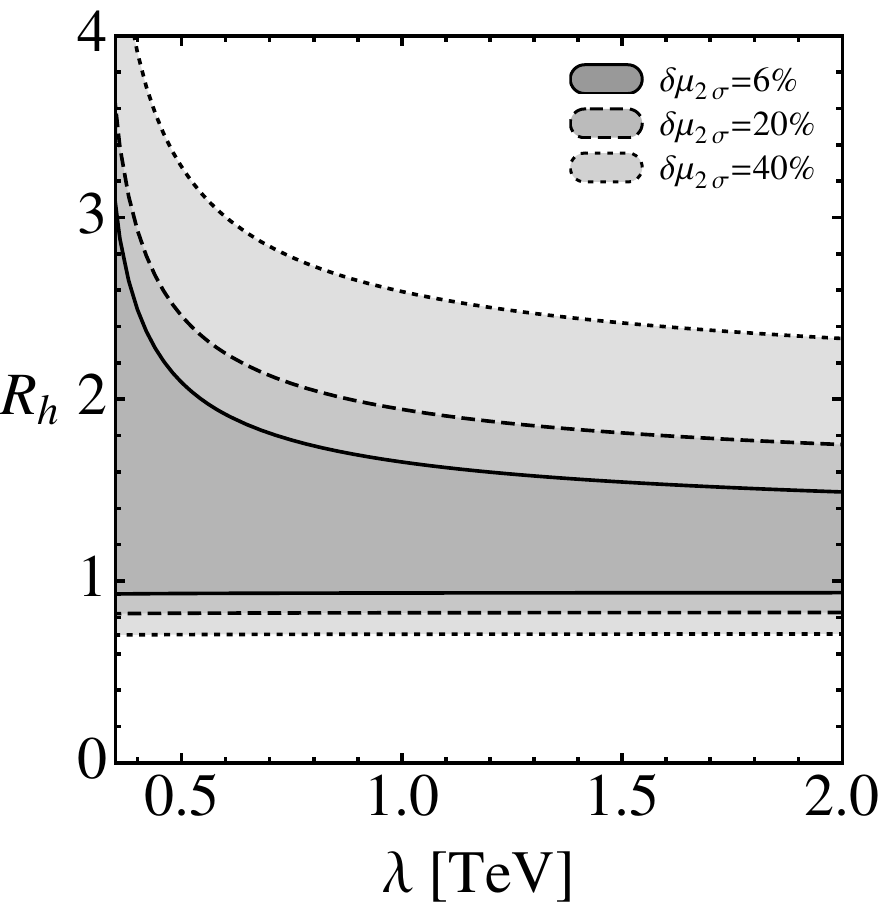} \qquad \includegraphics[height=2.6in]{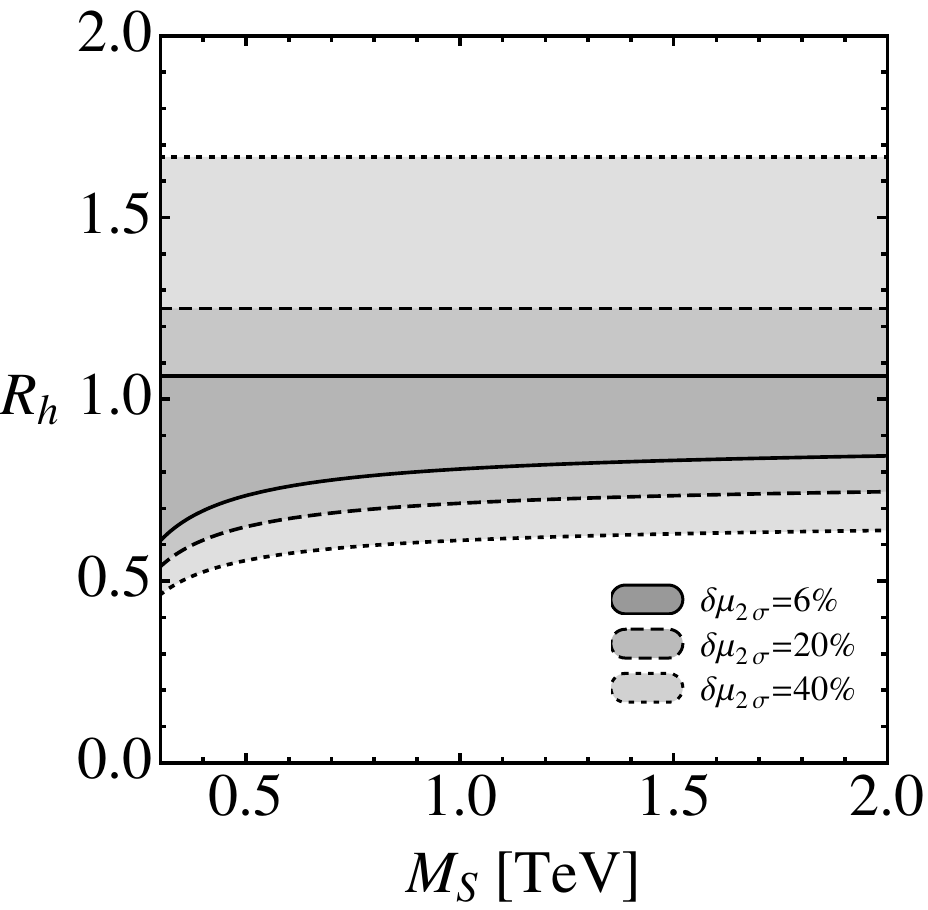}
\caption{Combined $2 \sigma$ LEP and future LHC WBF (or current all-channels) constraints on the Higgs width in the model with modified couplings and a UV cutoff (left) and a model with a mixed-in singlet scalar (right).}
\label{fig:widthconst}
\end{figure}

In \Fig{fig:widthconst} we show the expected $2 \sigma$ confidence contours on the total Higgs width that could be achieved with a future combination of WBF observations at the LHC with constraints from LEP for both the models of \Sec{sec:hwidth}.  Although it may be possible to achieve smaller uncertainties on the signal strength we will focus on the constraints determined from the $\delta \mu_{2 \sigma} = 40 \%$ band as this represents our most conservative estimate for the LHC at $300~\text{fb}^{-1}$ in the WBF channel.  

For $\lambda = 1$ TeV the $2 \sigma$ Higgs width constraints would be $0.71 \lesssim R_h \lesssim 2.59$.  The upper limit is much weaker than the lower limit because the precision electroweak constraints prefer increased Higgs couplings (see \Fig{fig:coupSTU}) and thus the combined constraint of \Eq{eq:comb} can tolerate significant increases in the Higgs width.  On the other hand, it is interesting that a strong lower limit can be placed on the Higgs total width.  For the model of a mixed-in singlet scalar the lower boundary of the constraint is similar to the left panel.  At $M_S = 1$ TeV the constraints are $0.61 \lesssim R_h \lesssim 1.67$.  

These potential future limits are quantitatively comparable to those obtained in \Sec{sec:univ} using current LHC data from global signal strength fits, however the purpose of the future constraints considered here is to reduce the model dependence of the constraint.  From this perspective this combination of past LEP constraints with future LHC Higgs measurements tailored to focus on $hVV$ couplings is very attractive as both constraints depend only on these couplings and the final constraint holds even under modifications in e.g.\  the $hGG$ coupling, which may occur due to new colored fields at the weak scale.

\section{Summary and Outlook}
\label{sec:conc}
In this paper, we have shown that including off-shell coupling measurements of the Higgs boson as performed by LEP adds important complementary information in the interpretation of LHC Higgs measurements. In particular we have shown that interpreting LEP results in terms of an $hVV$ coupling rescaling, we can understand LEP data as a Higgs coupling measurement. Accordingly we can break the degeneracy of the signal strength constraints similar to \cite{atlas,Khachatryan:2014iha,Caola:2013yja} to formulate a constraint on the total Higgs width. Bearing in mind certain theoretical issues that can arise if new physics does not follow a SM-pattern \cite{offshella,offshellb}, we have categorized the constraints into groups with varying degree of model-independence.  We have also discussed the theoretical limitations of this indirect constraint in \Sec{sec:LEP}.

We find that, assuming no new physics contributions to the $hGG$ coupling up to a scale of at least 1 TeV, the Higgs width is constrained to $0.73 \lesssim R_h (\lambda=1 \text{ TeV}) \lesssim 1.87$, based on the combination of Higgs-coupling measurements from LEP and the Higgs signal strength measurement from LHC8.  There is important model-dependence in this particular constraint as it assumes all Higgs couplings are rescaled in the same way.  This theoretical shortcoming in setting limits on $\Gamma_h$ can be avoided by considering fully-correlated production and decay modes such as weak boson fusion. Assuming a SM coupling pattern (i.e.\ we explicitly ignore the possibility of momentum-dependent couplings from higher dimensional operators or the presence of light electroweak degrees of freedom) we can make a conservative estimate for a future limit on the Higgs width using the LEP+LHC combination of $0.71 \lesssim R_h \lesssim 2.59$.

We stress that this result, although competitive with the expectation at a future Lepton Collider, is impacted by the logarithmic sensitivity to UV scales and should not be compared to constraints from a model-independent Lepton Collider measurement of the $hZ$ cross section as a probe of the $hZZ$ coupling.

Both the ATLAS and CMS collaborations have developed extensive analyses based on individually modified Higgs couplings in a vast number of production and decay mechanisms.  We have focussed on specific channels in this work, however it is likely that a combined global fit which floats all Higgs couplings and the Higgs width as an independent parameter, and then combines all of the available Higgs data at the LHC in a joint likelihood with the LEP constraints on Higgs couplings would lead to the strongest constraints on the Higgs width as interpreted within the popular `$\kappa$-framework'.  Moving towards the theoretically more robust setting of a Higgs EFT framework, the LEP constraints at tree-level and at one-loop are already well known (see for example the early references in \cite{HIGGSEWP}).  Thus in this case these constraints could be combined with LHC constraints on the coefficients of higher dimension operators which modify Higgs couplings, and if the Higgs width was included as a free parameter then again a strong constraint on the Higgs width would again result.  Such a constraint would fully demonstrate the power in combining the diverse strengths of LEP precision electroweak probes of the Higgs sector with the direct observations of the Higgs at the LHC.

{\it{Acknowledgments.}} MM would like to thank Jure Zupan for insightful comments in the initial stages of this work.  CE is supported in part by the IPPP Associateship programme.  MM acknowledges support from a CERN COFUND Fellowship.
This research was supported in part by the European Commission through the ``HiggsTools'' Initial
Training Network PITN-GA-2012-316704.



\begin{thebibliography}{99}

\bibitem{orig} 
  F.~Englert and R.~Brout,
  Phys.\ Rev.\ Lett.\  {\bf 13} (1964) 321.
  P.~W.~Higgs,
  Phys.\ Lett.\  {\bf 12} (1964) 132 and
  Phys.\ Rev.\ Lett.\  {\bf 13} (1964) 508.
  G.~S.~Guralnik, C.~R.~Hagen and T.~W.~B.~Kibble,
  Phys.\ Rev.\ Lett.\  {\bf 13} (1964) 585.
  

\bibitem{Chatrchyan:2012ufa}
  S.~Chatrchyan {\it et al.}  [CMS Collaboration],
  Phys.\ Lett.\ B {\bf 716} (2012) 30.

\bibitem{Aad:2012tfa}
  G.~Aad {\it et al.}  [ATLAS Collaboration],
  Phys.\ Lett.\ B {\bf 716} (2012) 1.

\bibitem{atlas}
    G.~Aad {\it et al.}  [ATLAS Collaboration],
  arXiv:1503.01060 [hep-ex].

\bibitem{Khachatryan:2014iha} 
  V.~Khachatryan {\it et al.}  [CMS Collaboration],
  Phys.\ Lett.\ B {\bf 736}, 64 (2014)
  [arXiv:1405.3455 [hep-ex]].


  
\bibitem{ChrisandMichael} 
      C.~Englert and M.~Spannowsky,
  Phys.\ Rev.\ D {\bf 90}, no. 5, 053003 (2014)
  [arXiv:1405.0285 [hep-ph]].


\bibitem{offshella}
    C.~Englert, Y.~Soreq and M.~Spannowsky,
  arXiv:1410.5440 [hep-ph].

\bibitem{offshellb}
    H.~E.~Logan,
  arXiv:1412.7577 [hep-ph].


  \bibitem{Offshell} 
    B.~Coleppa, T.~Mandal and S.~Mitra,
  Phys.\ Rev.\ D {\bf 90}, no. 5, 055019 (2014)
  [arXiv:1401.4039 [hep-ph]].
    J.~S.~Gainer, J.~Lykken, K.~T.~Matchev, S.~Mrenna and M.~Park,
  arXiv:1403.4951 [hep-ph].
  M.~Ghezzi, G.~Passarino and S.~Uccirati,
  PoS LL {\bf 2014} (2014) 072
  [arXiv:1405.1925 [hep-ph]].
    G.~Cacciapaglia, A.~Deandrea, G.~Drieu La Rochelle and J.~B.~Flament,
  Phys.\ Rev.\ Lett.\  {\bf 113}, no. 20, 201802 (2014)
  [arXiv:1406.1757 [hep-ph]].
    A.~Azatov, C.~Grojean, A.~Paul and E.~Salvioni,
  arXiv:1406.6338 [hep-ph].
    M.~Buschmann, D.~Goncalves, S.~Kuttimalai, M.~Schonherr, F.~Krauss and T.~Plehn,
  arXiv:1410.5806 [hep-ph].

\bibitem{ilc}
  S.~Liebler, G.~Moortgat-Pick and G.~Weiglein,
  arXiv:1502.07970 [hep-ph].
  S.~Liebler,
  arXiv:1503.07830 [hep-ph].

  
  
\bibitem{Caola:2013yja} 
  F.~Caola and K.~Melnikov,
  Phys.\ Rev.\ D {\bf 88}, 054024 (2013)
  [arXiv:1307.4935 [hep-ph]].
  
\bibitem{Kauer:2013qba}
  N.~Kauer and G.~Passarino,
  JHEP {\bf 1208} (2012) 116;
  N.~Kauer,
  JHEP {\bf 1312} (2013) 082;
  N.~Kauer,
  Mod.\ Phys.\ Lett.\ A {\bf 28} (2013) 1330015.


\bibitem{ciaran}
  J.~M.~Campbell, R.~K.~Ellis and C.~Williams,
  JHEP {\bf 1404} (2014) 060;
  J.~M.~Campbell, R.~K.~Ellis and C.~Williams,
  Phys.\ Rev.\ D {\bf 89} (2014) 053011;
  J.~M.~Campbell, R.~K.~Ellis, E.~Furlan and R.~R\"ontsch,
  arXiv:1409.1897 [hep-ph].

  \bibitem{Baak:2014ora} 
  M.~Baak {\it et al.}  [Gfitter Group Collaboration],
  Eur.\ Phys.\ J.\ C {\bf 74}, no. 9, 3046 (2014)
  [arXiv:1407.3792 [hep-ph]].

  \bibitem{Passarino:2010qk} 
  G.~Passarino, C.~Sturm and S.~Uccirati,
  Nucl.\ Phys.\ B {\bf 834}, 77 (2010)
  [arXiv:1001.3360 [hep-ph]].
  S.~Goria, G.~Passarino and D.~Rosco,
  Nucl.\ Phys.\ B {\bf 864} (2012) 530
  [arXiv:1112.5517 [hep-ph]].


\bibitem{Lopez-Val:2013yba} 
  D.~Lopez-Val, T.~Plehn and M.~Rauch,
  JHEP {\bf 1310}, 134 (2013)
  [arXiv:1308.1979 [hep-ph]].

\bibitem{Kribs:2007nz} 
  G.~D.~Kribs, T.~Plehn, M.~Spannowsky and T.~M.~P.~Tait,
  Phys.\ Rev.\ D {\bf 76}, 075016 (2007)
  [arXiv:0706.3718 [hep-ph]];
  B.~Holdom, W.~S.~Hou, T.~Hurth, M.~L.~Mangano, S.~Sultansoy and G.~Unel,
  PMC Phys.\ A {\bf 3}, 4 (2009)
  [arXiv:0904.4698 [hep-ph]].

\bibitem{sven}
  M.~Duhrssen, S.~Heinemeyer, H.~Logan, D.~Rainwater, G.~Weiglein and D.~Zeppenfeld,
  Phys.\ Rev.\ D {\bf 70} (2004) 113009
  [hep-ph/0406323].
  P.~Bechtle, S.~Heinemeyer, O.~Stal, T.~Stefaniak and G.~Weiglein,
  arXiv:1403.1582 [hep-ph].
  J.~Ellis, V.~Sanz and T.~You,
  JHEP {\bf 1407} (2014) 036
  [arXiv:1404.3667 [hep-ph]].
  C.~Englert, A.~Freitas, M.~M.~Mühlleitner, T.~Plehn, M.~Rauch, M.~Spira and K.~Walz,
  J.\ Phys.\ G {\bf 41} (2014) 113001
  [arXiv:1403.7191 [hep-ph]].
  J.~Ellis, V.~Sanz and T.~You,
  arXiv:1410.7703 [hep-ph].
  
  \bibitem{dob}
    B.~A.~Dobrescu and J.~D.~Lykken,
  JHEP {\bf 1302} (2013) 073
  [arXiv:1210.3342 [hep-ph]].

\bibitem{Espinosa:2012vu}
  J.~R.~Espinosa, M.~Muhlleitner, C.~Grojean and M.~Trott,
  JHEP {\bf 1209} (2012) 126
  [arXiv:1205.6790 [hep-ph]].



\bibitem{EFTJUST} 
  J.~R.~Espinosa, C.~Grojean, M.~Muhlleitner and M.~Trott,
  JHEP {\bf 1212}, 045 (2012)
  [arXiv:1207.1717 [hep-ph]].

\bibitem{HIGGSEWP} 
  S.~Alam, S.~Dawson and R.~Szalapski,
  Phys.\ Rev.\ D {\bf 57}, 1577 (1998)
  [hep-ph/9706542].
  R.~Barbieri, A.~Pomarol, R.~Rattazzi and A.~Strumia,
  Nucl.\ Phys.\ B {\bf 703}, 127 (2004)
  [hep-ph/0405040].
  R.~Barbieri, B.~Bellazzini, V.~S.~Rychkov and A.~Varagnolo,
  Phys.\ Rev.\ D {\bf 76}, 115008 (2007)
  [arXiv:0706.0432 [hep-ph]].
  R.~Contino,
  arXiv:1005.4269 [hep-ph].
  M.~Farina, C.~Grojean and E.~Salvioni,
  JHEP {\bf 1207}, 012 (2012)
  [arXiv:1205.0011 [hep-ph]].
  O.~Eberhardt, G.~Herbert, H.~Lacker, A.~Lenz, A.~Menzel, U.~Nierste and M.~Wiebusch,
  Phys.\ Rev.\ Lett.\  {\bf 109}, 241802 (2012)
  [arXiv:1209.1101 [hep-ph]].
  B.~Batell, S.~Gori and L.~T.~Wang,
  JHEP {\bf 1301}, 139 (2013)
  [arXiv:1209.6382 [hep-ph]].
  T.~Corbett, O.~J.~P.~Eboli, J.~Gonzalez-Fraile and M.~C.~Gonzalez-Garcia,
  Phys.\ Rev.\ D {\bf 87}, 015022 (2013)
  [arXiv:1211.4580 [hep-ph]].
  A.~Falkowski, F.~Riva and A.~Urbano,
  JHEP {\bf 1311}, 111 (2013)
  [arXiv:1303.1812 [hep-ph]].
    M.~Ciuchini, E.~Franco, S.~Mishima and L.~Silvestrini,
  JHEP {\bf 1308}, 106 (2013)
  [arXiv:1306.4644 [hep-ph]].
  
\bibitem{portal}
  T.~Binoth and J.~J.~van der Bij,
  Z.\ Phys.\ C {\bf 75} (1997) 17
  [hep-ph/9608245].
  R.~Schabinger and J.~D.~Wells,
  Phys.\ Rev.\ D {\bf 72} (2005) 093007
  [hep-ph/0509209].
  B.~Patt and F.~Wilczek,
  hep-ph/0605188.
  C.~Englert, T.~Plehn, D.~Zerwas and P.~M.~Zerwas,
  Phys.\ Lett.\ B {\bf 703} (2011) 298
  [arXiv:1106.3097 [hep-ph]].
  D.~Bertolini and M.~McCullough,
  JHEP {\bf 1212}, 118 (2012)
  [arXiv:1207.4209 [hep-ph]].

  \bibitem{Baer:2013cma} 
  H.~Baer, T.~Barklow, K.~Fujii, Y.~Gao, A.~Hoang, S.~Kanemura, J.~List and H.~E.~Logan {\it et al.},
  arXiv:1306.6352 [hep-ph].
  
  \bibitem{Accomando:2004sz} 
  E.~Accomando {\it et al.}  [CLIC Physics Working Group Collaboration],
  hep-ph/0412251.
  
  \bibitem{Gomez-Ceballos:2013zzn} 
  M.~Bicer {\it et al.}  [TLEP Design Study Working Group Collaboration],
  JHEP {\bf 1401}, 164 (2014)
  [arXiv:1308.6176 [hep-ex]].
  
\bibitem{cepc} 
   See \url{http://cepc.ihep.ac.cn}.
  
  \bibitem{Peskin:1990zt} 
  M.~E.~Peskin and T.~Takeuchi,
  Phys.\ Rev.\ Lett.\  {\bf 65}, 964 (1990).
  M.~E.~Peskin and T.~Takeuchi,
  Phys.\ Rev.\ D {\bf 46}, 381 (1992).
  
\bibitem{Duehrssen} 
  M.~Duehrssen,
  Moriond Talk, 2015

\bibitem{Khachatryan:2014jba}
  V.~Khachatryan {\it et al.}  [CMS Collaboration],
  arXiv:1412.8662 [hep-ex],
  see \url{https://twiki.cern.ch/twiki/bin/view/CMSPublic/Hig14009TWiki}.

\bibitem{vbfellis}
J.~M.~Campbell and R.~K.~Ellis,
  arXiv:1502.02990 [hep-ph].

\bibitem{vbfnlo}
J.~Baglio, J.~Bellm, F.~Campanario, B.~Feigl, J.~Frank, T.~Figy, M.~Kerner and L.~D.~Ninh {\it et al.},
  arXiv:1404.3940 [hep-ph].
  K.~Arnold, M.~Bahr, G.~Bozzi, F.~Campanario, C.~Englert, T.~Figy, N.~Greiner and C.~Hackstein {\it et al.},
  Comput.\ Phys.\ Commun.\  {\bf 180} (2009) 1661
  [arXiv:0811.4559 [hep-ph]].


\bibitem{wbfscales}
  B.~Jager, C.~Oleari and D.~Zeppenfeld,
  JHEP {\bf 0607} (2006) 015
  [hep-ph/0603177].
  G.~Bozzi, B.~Jager, C.~Oleari and D.~Zeppenfeld,
  Phys.\ Rev.\ D {\bf 75} (2007) 073004
  [hep-ph/0701105].


\bibitem{Bahr:2008pv}
  M.~Bahr, S.~Gieseke, M.~A.~Gigg, D.~Grellscheid, K.~Hamilton, O.~Latunde-Dada, S.~Platzer and P.~Richardson {\it et al.},
  Eur.\ Phys.\ J.\ C {\bf 58} (2008) 639
  [arXiv:0803.0883 [hep-ph]].

\bibitem{Alwall:2011uj} 
  J.~Alwall, M.~Herquet, F.~Maltoni, O.~Mattelaer and T.~Stelzer,
  JHEP {\bf 1106}, 128 (2011)
  [arXiv:1106.0522 [hep-ph]].
  
\bibitem{alpgen}
  M.~L.~Mangano, M.~Moretti, F.~Piccinini, R.~Pittau and A.~D.~Polosa,
  JHEP {\bf 0307} (2003) 001
  [hep-ph/0206293].

\bibitem{detector}
ATLAS collaboration, ATL-PHYS-PUB-2013-004.


\bibitem{Cacciari:2008gp} 
  M.~Cacciari, G.~P.~Salam and G.~Soyez,
  JHEP {\bf 0804}, 063 (2008)
  [arXiv:0802.1189 [hep-ph]].

\bibitem{Kauer:2000hi}
  N.~Kauer, T.~Plehn, D.~L.~Rainwater and D.~Zeppenfeld,
  Phys.\ Lett.\ B {\bf 503} (2001) 113
  [hep-ph/0012351].

\bibitem{Barger:1994zq}
  V.~D.~Barger, R.~J.~N.~Phillips and D.~Zeppenfeld,
  Phys.\ Lett.\ B {\bf 346} (1995) 106
  [hep-ph/9412276].


\end{thebibliography}
\end{document}